# Optical and electronic properties of two dimensional graphitic silicon carbide

Xiao Lin[1,2], Yang Xu[1], Shisheng Lin[1,3,a)], Ayaz Ali Hakro[1], Te Cao[1], Hongsheng Chen[1,2,b)], and Baile Zhang[4]

[1]Department of Information Science and Electronic Engineering
Zhejiang University, Hangzhou 310027, China
[2]The Electromagnetics Academy, Zhejiang University, Hangzhou 310027, China
[3]State Key Laboratory of Modern Optical Instrumentation
Zhejiang University, Department of optical engineering, Hangzhou 310027, China
[4]Singapore-MIT Alliance for Research and Technology
(SMART) Centre, Singapore 117543, Singapore

**Abstract:**

Optical and electronic properties of two dimensional few layers graphitic silicon carbide (GSiC), in particular monolayer and bilayer, are investigated by density functional theory and found different from that of graphene and silicene. Monolayer GSiC has direct bandgap while few layers exhibit indirect bandgap. The bandgap of monolayer GSiC can be tuned by an in-plane strain. Properties of bilayer GSiC are extremely sensitive to the interlayer distance. These predictions promise that monolayer GSiC could be a remarkable candidate for novel type of light-emitting diodes utilizing its unique optical properties distinct from graphene, silicene and few layers GSiC.

---

[a] shisheng.daniel.lin@gmail.com (S.S. LIN)

[b] hansomchen@zju.edu.cn (H. S. Chen)

Atomic thin materials, especially graphene, have intrigued lots of attentions in the area of condensed matter physics, materials science, and electronic engineering. The two dimensional (2D) massless Dirac Fermions in graphene permit a few extraordinary phenomena, such as abnormal quantum Hall effect and fine constant related universal absorption, which have rarely been observed in other condensed materials.[1-4] The carrier mobility in graphene can be as high as 20 $m^2$/Vs,[5] which is the record of all existing materials and promises the application of high-frequency electronic devices up to several hundred GHz.[6,7] Silicene, the silicon counterpart of graphene, has also received worldwide attentions.[8-12]

The main drawback of graphene and silicene is that they have a zero electronic bandgap. For optoelectronic applications, such as light-emitting diodes (LEDs) or solar cells, the active material should have an approximate bandgap. In addition, a large exciton binding energy should be desirable for a high internal quantum efficiency. Graphitic monolayer silicon carbide (SiC), a honeycomb stacked by silicon and carbon in sequence, is such a material which uniquely combines a large bandgap and a very high exciton binding energy up to 2.0 eV.[13] Such a high exciton energy may even promise Bose–Einstein condensate (BEC) effect, in analogy with $Cu_2O$.[14]

Traditionally, SiC is sphalerite or wurtzite crystal with indirect bandgaps, and has been widely used in high temperature, power and frequency devices.[15] Calculations have predicted that SiC could be graphitic structure when it comes to a few layers thickness.[16-20] Graphitic silicon carbide (GSiC)[12,21,22,23] has been produced in recent experiments,[24,25] where GSiC shows a better photoluminescence (PL) ability than its sphalerite or wurtzite counterpart.[25] For advancing the research on GSiC, a detailed comparison between monolayer GSiC and few layers GSiC, as well as graphene and silicene, is highly desirable to fully resolve its fundamental physical interests and potential applications.

Herein we report the optical and electronic properties of layered 2D GSiC, in particular for monolayer and bilayer, by density functional theory. We find that $sp^2$-bonded layered GSiC could transform from indirect into direct bandgap material gradually when the thickness becomes thinner and finally a direct bandgap can be obtained in monolayer GSiC, which can be further tuned by in-plane strain.

As shown in Fig. 1a, the band structure of monolayer GSiC is presented along with that of graphene and silicene. For graphene and silicene, Dirac point is found at symmetric $K$ point in the hexagonal Brillouin zone.[10,12,26,27] In contrast, monolayer GSiC is a binary compound and has a direct bandgap of ~2.5 eV due to the ionicity.[21,22] The optical properties of graphene, silicene and monolayer GSiC are shown in Fig. 1b. Silicene and graphene have universal optical conductivity below 1.0 eV and visible photon energy, respectively. While monolayer GSiC absorbs photons at visible light range, there is no universal optical conductivity. Under normal-incidence, graphene can absorb 17% light at 4.2 eV and monolayer GSiC can absorb 21% light at 3.3 eV. In low-buckled silicene, however, the strong interaction between the two planes gives rise to two prominent absorption peaks: 13% absorption at 1.8 eV and 37% absorption at 4.0 eV. Meanwhile, it makes the band structure around $M$ and $\Gamma$ points of silicene more complicated than that of graphene and monolayer GSiC. For those materials, all these prominent peaks are mainly induced by the electron transition around *M* point, whereas some small peaks in silicene are induced by the transition around *Γ* point. It is worthy noting that negative real part of permittivity could be found in all those materials at ultraviolet range while silicene also exhibits negative permittivity at visible range. A negative permittivity is a favored figure of merit in plasmonics and metamaterials.[28,29] Therefore, monolayer GSiC possesses properties different from graphene and silicene, which should enable novel photonic and electronic applications.[11,25]

As shown in Fig. 2, through changing the bond length, the dependence of optical and electronic properties of monolayer GSiC on the uniform in-plane strain is studied. We note that the theoretically optimized bond length (1.79 Å) is larger than experimental value (1.54 Å),[25] which could be possibly caused by defects. Owing to the ionicity in monolayer GSiC, the conduction band at *K* and *M* points mainly consists of one $\pi^*$ band arising from Si $3p_z$ orbitals.[13] As shown in Fig. 2b, decreasing bond length enhances repulsive force between neighboring Si and C atoms, resulting in the linear increase of the direct bandgap at *K* point ($\Delta_{KK}$) and indirect bandgap between *K-M* points ($\Delta_{KM}$). Moreover, decreasing bond length also promotes the overlap of neighboring Si $3p_z$ orbitals, leading to that $\Delta_{KM}$ changes slower than $\Delta_{KK}$. As seen from Fig. 2, monolayer GSiC would transit from direct into indirect bandgap semiconductor (namely $\Delta_{KK} > \Delta_{KM}$) when bond length is less than 1.74 Å. Hence in our previous experiment, monolayer GSiC with 1.54 Å bond length should own indirect bandgap, and the PL of monolayer GSiC might be further enhanced by elongating the bond length.[25] As the bond length varies within the range of 1.74 to 2.0 Å, monolayer GSiC has a direct bandgap ($\Delta_{KK} < \Delta_{KM}$). On the other hand, when the bond length increases above 2.0 Å (Fig. 2a), the conduction band minimum at $\Gamma$ point becomes smaller than that at $K$ point and monolayer GSiC would transit into indirect bandgap material. Fig. 2c shows the optical conductivity spectrum of monolayer GSiC, and the maximum absorption peak enhances its intensity and blue-shifts when the bond length decreases. Our results show that strain could be an effective way to realize desired functionality in monolayer GSiC.

The band structure of GSiC is sensitive to the layer number (Fig. 3). Remarkably, the monolayer GSiC has a direct bandgap while few layers GSiC sheets have an indirect bandgap. Our observation is similar to the case of transition metal dichalcogenides $MoS_2$. Since Mo atoms locate at the middle of the S-Mo-S unit cell in $MoS_2$, the interlayer coupling has little influence on the direct bandgap at $K$ point, which is induced by strongly localized $d$ orbitals at Mo atom site.[30] However, the interlayer coupling has strong influence on the indirect bandgap away from the $K$ point, which originates from a linear combination of $d$ orbitals on Mo atoms and antibonding $p_z$ orbitals on S atoms.[30] Hence the few layers $MoS_2$ belong to indirect bandgap materials originated from interlayer coupling while monolayer $MoS_2$ has a direct bandgap.[30] As an $sp$-bonded system of monolayer GSiC, the conduction band minimum is relatively flat along the $K$-$M$ direction and $K$ point locates lower than $M$ point (Fig. 3). The interlayer coupling in few layers GSiC, which is caused by the overlap of Si $3p_z$ orbitals with the right facing C $2p_z$ orbitals in other layers, has a strong impact on the conduction band around *M* and *K* points. As illustrated in Fig. 3, when compared with monolayer, the conduction band minimum at *M* point in few layers GSiC is much lowered, while that at *K* point only transversely slightly moved away from the original *K* point. Hence, all few layers GSiC transit into indirect bandgap semiconductors as conduction band minimum around *K* point becomes higher than that at *M* point. As the lattice plane interacts only with the nearest two layers in GSiC, the band structures change dramatically from monolayer to trilayer and change slightly from quadrilayer to hexalayer. Deviating from the *d*-electron materials (such as $MoS_2$), our results disclose a new *sp*-bonded materials system where electronic band structures can be controlled by layer thickness.

Different from monolayer, three prominent absorption peaks, locating at ~2.4, 3.3 and 3.4 eV, appear in bilayer GSiC with an interlayer distance of 3.47 Å (Fig. 4). However, as the data of interlayer distance of ultrathin GSiC are scattering,[24,25] we explore the optical and electronic properties of bilayer GSiC as a function of interlayer distance. The second and third absorption peaks in bilayer GSiC, which are mainly caused by electron transitions at *M* point (denoted as $\Delta_{M1}$ and $\Delta_{M2}$ in Fig. 4a), could blue-shift with smaller interlayer distance. In addition, when the interlayer distance decreases at the range of 3.0-3.8 Å, the direct bandgap monotonously decreases at *M* point ($\Delta_M$) and increases near *K* point ($\Delta_K$), respectively. As a result, the bilayer GSiC initiates photonic absorption in lower photon energy and the first absorption peak is greatly enhanced with slight blue-shift (Fig. 4c). Hence, the interlayer distance has a great impact on the optical and electronic properties of bilayer GSiC. The bandgap in bilayer GSiC is always indirect regardless of the interlayer distance, which implies monolayer GSiC is superior to few layers GSiC in practical LEDs.

In summary, we have investigated the optical and electronic properties of GSiC as a function of bond length and interlayer distance. Layered 2D GSiC reveals one prominent advantage, namely such capability of an indirect to direct bandgap transition from few-layer to monolayer. By applying strain to monolayer, a tunable bandgap can be reached, making monolayer GSiC suitable for the LEDs and solar cells. Our theoretical investigation is expected to stimulate and guide the design or characterization of photonic devices based on layered GSiC.

## Acknowledgements

The authors would like to thank Shanghai Supercomputer Center for simulation support. This work is sponsored by the NSFC (Grants Nos. 60801005, 60990320, 60990322, and 61006077), the Foundation for the Author of National Excellent Doctoral Dissertation of PR China (Grant No. 200950), the Zijin Award at ZJU, Fundamental Research Funds for the Central Universities, and SRFDP (No. 20100101120045). S. S. LIN thanks to funding support from State Key Laboratory of Modern Optical Instrumentation (111306*A61001) and the China postdoctoral science foundation (20100480083 and 201104714).

# Figures

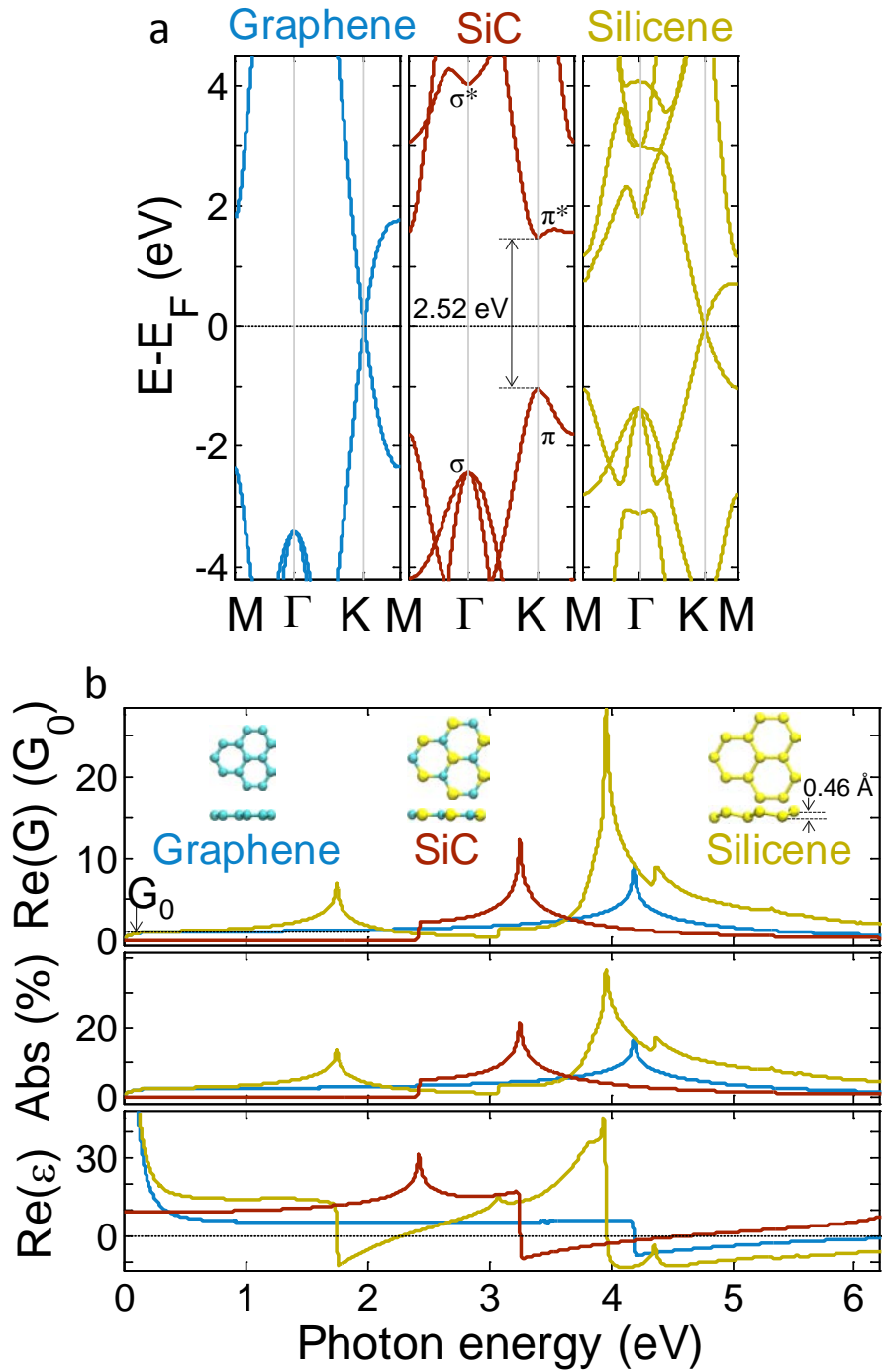

**Figure 1:** Comparison of (a) band structure, (b) optical conductivity, relative permittivity (real part, Re(G) and Re(ε)), and normal-incidence absorption of monolayer GSiC, graphene, and silicene. $E_F$ is Fermi energy. Insets in (b) are schematics of atomic structures with top and side views, where silicene is low-buckled. Dirac point and universal optical conductivity ($G_0$, dashed black line in optical conductivity spectra) are found in graphene and silicene, however not in GSiC due to the ionicity.

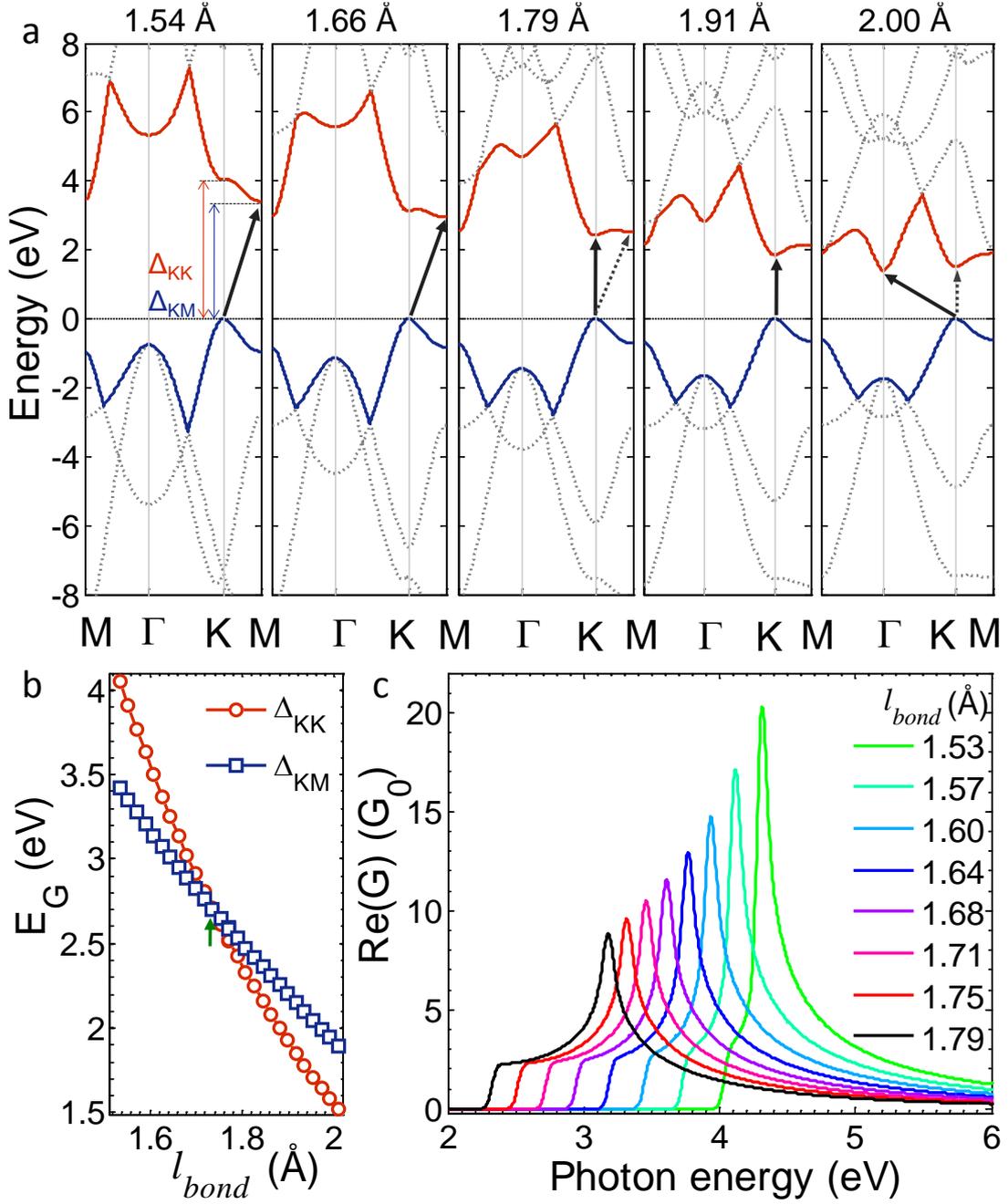

**Figure 2:** (a) Band structures of monolayer GSiC with various bond lengths ($l_{bond}$). The zero of energy is set to the top of valence band at *K* point. (b) Variation in Energy gap ($E_G$) of $\Delta_{KK}$ and $\Delta_{MM}$ (denoted as red and blue arrows in (a)). From (a) and (b), the direct bandgap can be tuned to indirect one by varying $l_{bond}$ shorter than 1.74 Å (marked by green arrow in (b)) or longer than 2.0 Å. (c) Dependence of optical properties on $l_{bond}$. Optical conductivity spectrum blue-shifts by shortening $l_{bond}$.

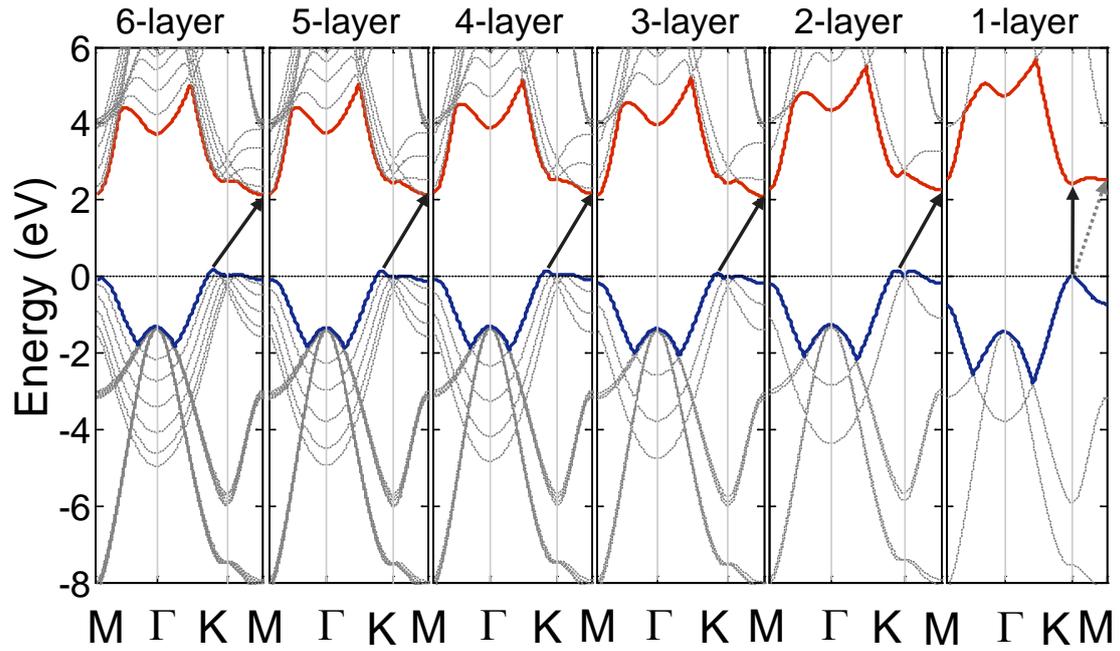

**Figure 3:** Layer numbers dependent band structure of GSiC. The zero energy is set to the valence top at K point. The indirect bandgap transits to direct one when few layers GSiC thinned to monolayer.

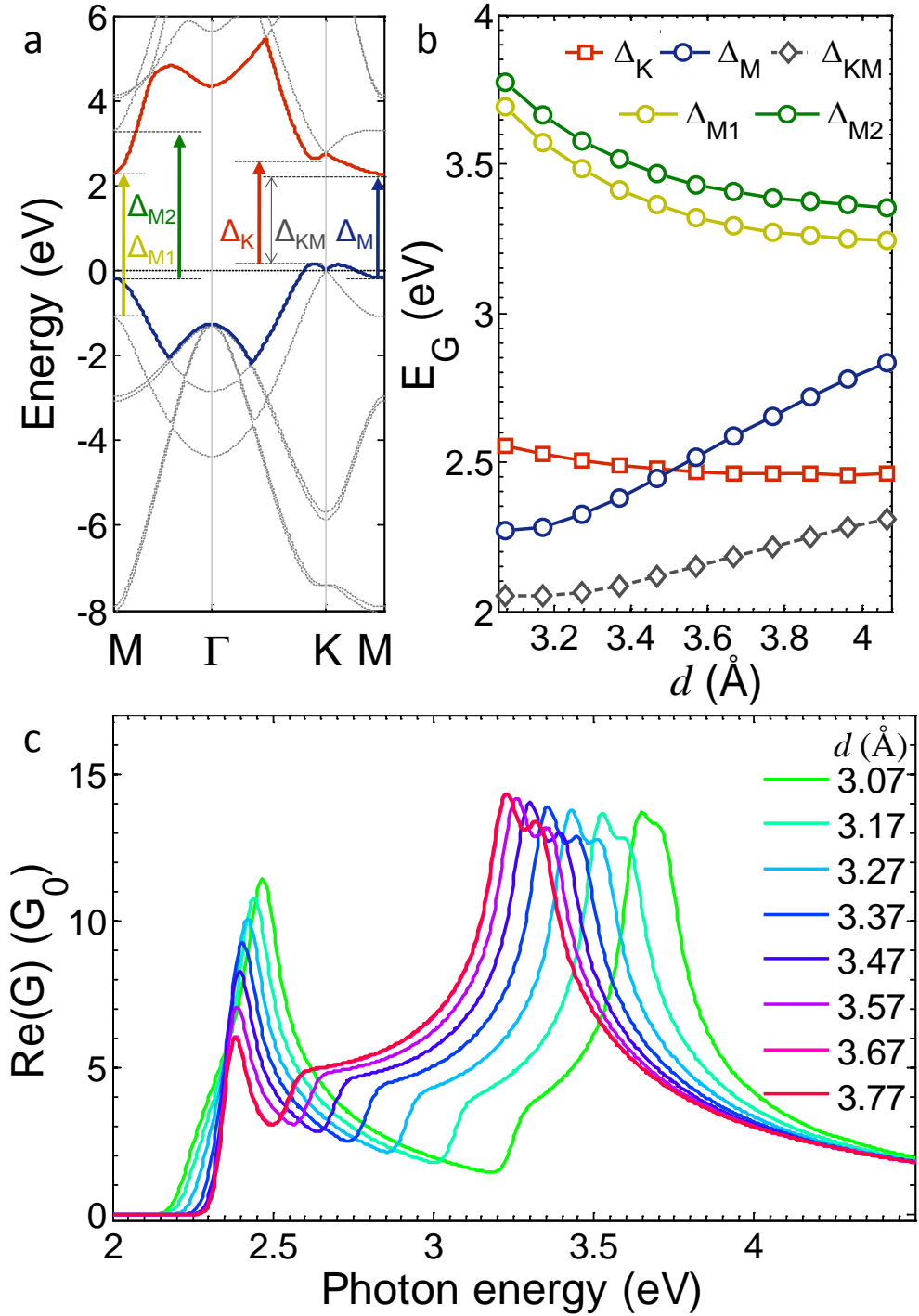

**Figure 4:** Interlayer distance (*d*) dependent (a) band structure, (b) different energy gaps (E$_G$) and (c) optical properties of bilayer GSiC. Indirect Δ$_{KM}$ (denoted as black arrow in (a)) is smallest, making bilayer GSiC always an indirect bandgap semiconductor regardless of *d*. When decreasing *d*, direct Δ$_K$, Δ$_{M1}$, and Δ$_{M2}$ (red, yellow, and green arrows in (a), respectively) increase and three peaks in Re(G)

spectra all blue-shift, while direct $\Delta_M$ (blue arrow in (a)) decreases and photon absorption initiates at lower energy in (c).

# Supporting Information

Xiao Lin, Yang Xu, Shisheng Lin, Ayaz Ali Hakro, Te Cao, Hongsheng Chen, and Baile Zhang

Layered GSiC are studied with density functional theory in generalized gradient approximation (GGA) by optical package of SIESTA codes,[S1] in which double ζ polarized numerical atomic basis sets and norm-conserving pseudo-potentials are used. The exchange-correlation function of GGA is represented by Revised Perdew-Burke-Ernzerh (RPBE) approximation.[S2] A vacuum space (≥2.5 nm) between periodic images of the slabs is applied in order to avoid spurious interaction. A k-grid-Monkhorst-Pack mesh of 100 × 100 × 1 and a 300 Ry energy cutoff are set to ensure converged GGA results. Optical properties are calculated at 300 K electronic temperature and with light incidence perpendicular to the 2D plane throughout this paper, where electric and magnetic fields are parallel to GSiC plane. To obtain smooth optical spectra, when Gaussian broadening is set to be 0.04 eV, 600 × 600 × 9 number of optical mesh points are needed. In order to fulfill the *f*-sum rule, the energy range sets to be 0-40 eV. When obtaining the imaginary part of relative permittivity from SIESTA, its real part can be derived by applying the Kramers-Kronig relations[S3] $\varepsilon_R(\omega) - \varepsilon_\infty = \frac{1}{\pi} PV \int_{-\infty}^{\infty} d\alpha \frac{\varepsilon_I(\alpha)}{\alpha - \omega}$. One can crossly check the quality of the Kramers-Kronig transformation by deriving the imaginary part with $\varepsilon_I(\omega) = -\frac{1}{\pi} PV \int_{-\infty}^{\infty} d\alpha \frac{\varepsilon_R(\alpha) - \varepsilon_\infty}{\alpha - \omega}$. The relation of the complex relative permittivity $\bar{\varepsilon}$ and optical conductivity $\bar{G}$ is $\bar{\varepsilon} = 1 + i\frac{\bar{G}}{d\omega\varepsilon_0}$ by using Maxwell equations,[S3] where $d$ is thickness of the studied monolayer or few layers 2D materials and ω is photon frequency. Hence, the complex $\overline{G} = \text{Re}(\overline{G}) + i\,\text{Im}(\overline{G})$ can be calculated.

In our calculation, the optimized lattice constant of monolayer GSiC is 3.10 Å as shown in Fig. S1, in agreement with previous theory,[12,13,22,23] and the lattice constant of graphene and silicene are 2.46 and 3.866 Å,[10] respectively. The thicknesses of graphene, silicene and monolayer GSiC are 3.35, 3.7, and 3.47 Å from experiments,[8,25,S4] respectively, which are used for the calculation of normal-incidence absorption based on Fresnel theory.[S3] For the completeness of research in Fig. 1, more optical parameters of monolayer GSiC, graphene, and silicene are also shown in Fig. S2, where the differences in these three 2D materials can be clearly seen. For the completeness of research in Fig. 2 and Fig. 4, the real part of permittivity can be tuned by applying in-plane strain in monolayer GSiC or by varying the interlayer distance in bilayer GSiC as shown in Fig. S3. Hence tunable plasmonics might be designed based on GSiC.

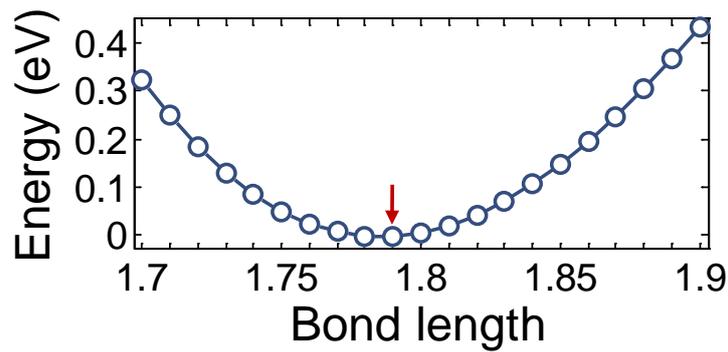

**Figure S1:** Optimization of bond length in monolayer GSiC. The optimized bond length is 1.79 Å, denoted as the red arrow.

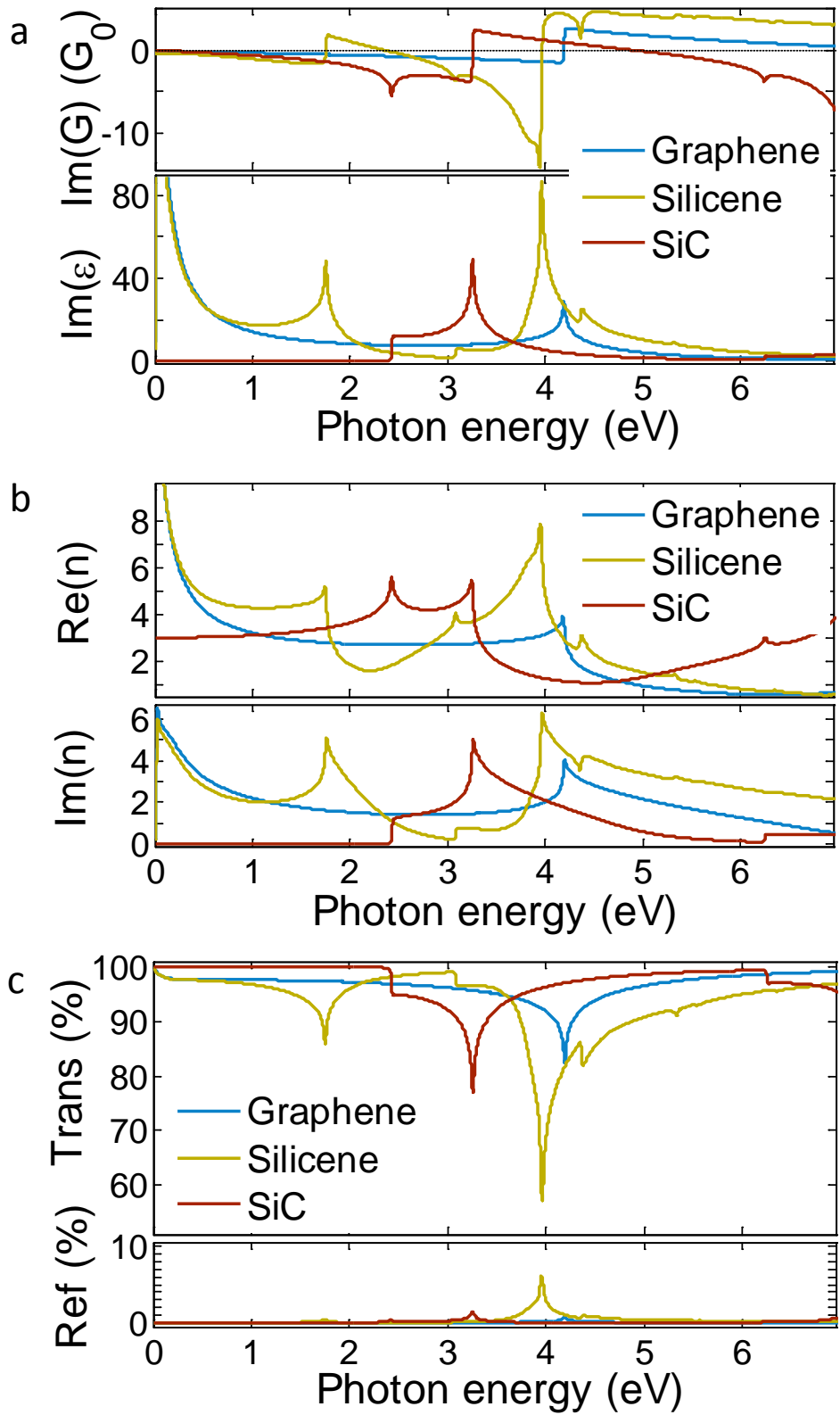

**Figure S2:** Different optical properties in monolayer GSiC, graphene, and silicene. (a) optical conductivity and relative permittivity (imaginary part, Im(G) and Im(ε)). (b) complex refractive index. (c) normal-incidence reflection and transmission.

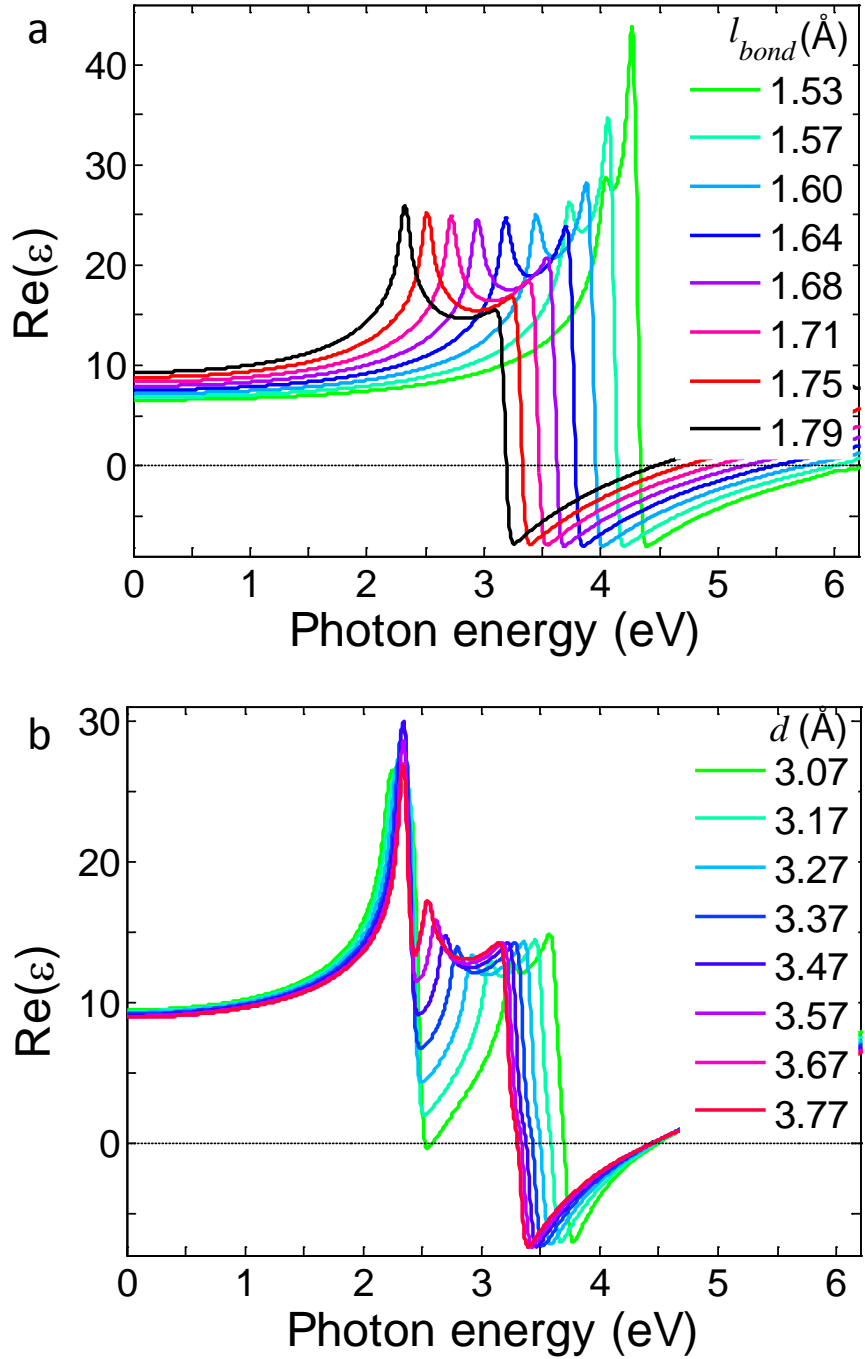

**Figure S3:** (a) Real part of permittivity (Re(ε)) in monolayer GSiC is sensitive to bond length ($l_{bond}$). The frequency range of negative Re(ε) blue-shifts by shortening $l_{bond}$. (b) Re(ε) in bilayer GSiC can be tuned by varying interlayer distance (*d*). The near-ultraviolet negative Re(ε) range blue-shifts by decreasing *d*. Moreover, one more visible negative Re(ε) range could emerge when *d* is less than 3.07 Å.

## Supporting References